\documentclass[journal,onecolumn]{IEEEtran}

\usepackage{epsfig}
\usepackage{graphicx}
\usepackage{graphics}
\usepackage{epstopdf} 
\usepackage{tikz}
\usetikzlibrary{calc,shapes.geometric}
\usetikzlibrary{arrows,snakes,backgrounds}
\usepackage{pgf, tikz}
\usetikzlibrary{arrows, automata, positioning}
\usetikzlibrary{quotes,angles}
\usepackage{amssymb}
\usepackage{amsmath}
\usepackage{xcolor}

\usepackage{tcolorbox}
\tcbset{colframe=black,colback=white,colupper=black,
fonttitle=\bfseries,nobeforeafter,center title,size=small,arc=0pt,outer arc=0pt}

\ifCLASSINFOpdf
\else
\fi

\def\nexto{\kern -0.54em}
\def\mean{{\rm {I\ \nexto E}}}
\def\prob{{\rm {I\ \nexto P}}}

\def\pfa{{\rm P_{FA}}}

\def\I{{\rm {I\ \nexto I}}}

\def\nexto{\kern -0.54em}
\def\mean{{\rm {I\ \nexto E}}}
\def\prob{{\rm {I\ \nexto P}}}

\begin{document}

\title{Armoured Fighting Vehicle Team Performance Prediction against Missile Attacks with Directed Energy Weapons}
\author{Graham  V. Weinberg and Mitchell M. Kracman\\
Graham.Weinberg@dst.defence.gov.au
 }
\maketitle

\markboth{AFV Defence: \today}%
{}

\begin{abstract}
A recent study has introduced a procedure to quantify the survivability of a team of armoured fighting vehicles when it is subjected to a single missile attack. In particular this study investigated the concept of collaborative active protection systems, focusing on the case where vehicle defence is provided by high power radio frequency directed energy weapons.
The purpose of the current paper is to demonstrate how this analysis can be extended to account for more than one missile threat. This is achieved by introducing a jump stochastic process whose states represent the number of missiles defeated at a given time instant. Analysis proceeds through consideration of the sojourn times of this stochastic process, and it is shown how consideration of these jump times can be related to transition probabilities of the auxiliary stochastic process. The latter probabilities are then related to the probabilities of detection and disruption of missile threats. The sum of these sojourn times can then be used to quantify the survivability of the  team at any given time instant. Due to the fact that there is much interest in the application of high energy lasers in the context of this paper, the numerical examples will thus focus on such directed energy weapons for armoured fighting vehicle  team defence. 
\end{abstract}


\setlength {\abovedisplayskip} {6pt plus 3.0pt minus 4.0pt}
\setlength {\belowdisplayskip} {6pt plus 3.0pt minus 4.0pt}

\section{Introduction}
\label{sec:1}
Performance prediction of collaborative active defence systems for a team of armoured fighting vehicles (AFV) is of significant importance to defence forces investing in modern technology. In particular, active protection systems (APS) are being examined in terms of their military utility \cite{feng}. The concept of collaboration with such systems, especially when coupled with emerging disruptive technology for threat defeat, is of considerable interest to Defence Science and Technology Group. An overview of APS can be found in \cite{meyer}, while a comprehesive report on it, presented to the United States Congress, can be found in \cite{usarmy}. 
Some commerically produced examples include the Raytheon APS \cite{apsref}, Rheinmetall's Active Defence System \cite{apsref2} and the Rafael Israeli Trophy system \cite{trophy}. 
An APS is integrated into the AFV and is designed to automatically detect and track incoming threats, and then provide a countermeasure to neutralise the threat \cite{usarmy}.
The countermeasure can be classified as either delivering a soft or hard kill \cite{meyer}. Soft kills result from countermeasures which either deceive or disrupt the missile's guidance and control, so that they are applying electronic warfare countermeasures. A hard kill results when the missile is disrupted or destroyed by direct weapons engagement, such as  reactive armour and anti-ballistic missiles. 

Toward providing this performance prediction capability for AFV defence through APS, a recent study examined an approach to model a team's survivability in the presence of a single missile threat \cite{weinberg21}. The latter study discretised time in order to facilitate the mathematical analysis, and provided some examples of performance prediction of a simplified version of collaborative APS (C-APS).  This system assumed that a subset of the AFVs is responsible for target detection and tracking, while another subset is assumed to be responsible for threat engagement. The system's {\em modus operandi} was that if a vehicle with detection capability sensed the target, it then would schedule the nearest vehicle with disruption capability, to counter the missile threat. A key feature of this study was to examine the military utility of high power radio frequency (HPRF) directed energy weapons (DEWs) 
\cite{nielsen}. 

The purpose of the current study is to demonstrate how the analysis in \cite{weinberg21} can be extended to the situation of multiple missile threats. In order to do this it is shown how the problem under consideration can be re-interpreted in terms of sojourn times of an auxiliary jump stochastic process, whose states are the number of missiles defeated at a given time. Due to this novel approach it is then possible to remove the limitation of discretised time. It will be shown that the case of a single missile threat is somewhat simple, while the case of two missiles is more complicated but produces an elegant solution.  Extensions beyond the two missile case will also be discussed. 

Due to the current interest in the application of high energy laser (HEL) DEWs for missile defence the examples investigated will apply such DEWs for the AFV team defence. This has necessitated the development of a new expression for the HEL effect on a target, which accounts for target dwell time. A novel way to account for dwell time is thus introduced, producing a new expression for HEL DEW impact on a target.

Towards these objectives the paper is structured as follows. Section \ref{sec:2} introduces the mathematical preliminaries and outlines the general approach. Section \ref{sec:3} then derives an explicit expression for team survivability in the presence of a single missile threat. Sections \ref{sec:4} demonstrates how to extend this to account for two missiles.  Finally, Section \ref{sec:5} specialises the general developments in Sections \ref{sec:3} and \ref{sec:4} to provide some tangible examples of performance, following the approach in \cite{weinberg21} but focusing on HEL DEWs for APS countermeasures.

\section{Mathematical Preliminaries}
\label{sec:2}
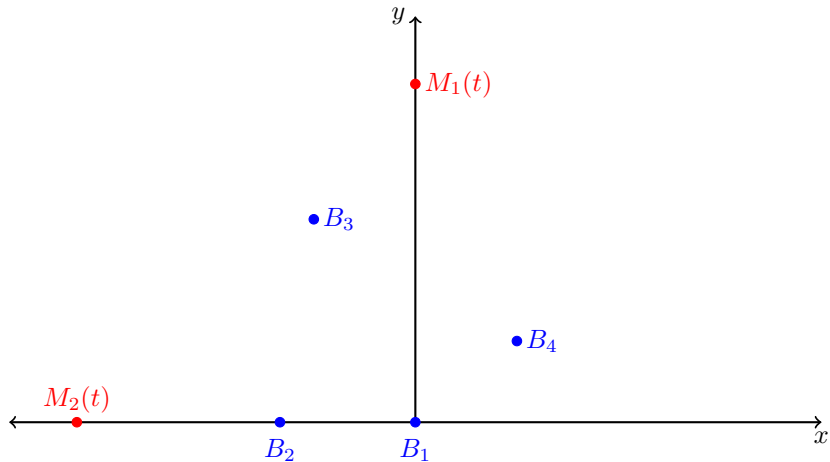
\begin{figure}[h]
\centering
\begin{tikzpicture}[scale=0.9]
   
    \draw[thick,<->] (-6,0)--(6,0) node[below] {$x$}; 
    \draw[thick,->] (0,0)--(0,6) node[left] {$y$};
   \draw[red,fill] coordinate (Mn) (0,5) circle (2pt) node[right]{$M_1(t)$}; 
    \draw[red,fill] coordinate (Mn2) (-5,0) circle (2pt) node[above]{$M_2(t)$}; 
    \draw[blue,fill] coordinate (B1) (0,0) circle (2pt) node[below =0.1cm]{$B_1$};   
    \draw[blue,fill] coordinate (B2)(-2,0) circle (2pt) node[below =0.1cm]{$B_2$};
    \draw[blue,fill] coordinate (B3)(-1.5,3) circle (2pt) node[right] {$B_3$};
    \draw[blue,fill] coordinate (B4) (1.5, 1.2) circle (2pt) node[right] {$B_4$};
\end{tikzpicture}
\caption{Geometry of the AFV combat scene under examination. In this example two missiles, denoted $M_1(t)$ and $M_2(t)$, have been fired toward the AFV team.
The latter, denoted $B_1, B_2, B_3$ and $B_4$, are assumed stationary during the engagement. In terms of defence, all team members will at some time be in a position to provide defence of $B_1$, which is targeted by $M_1(t)$. By contrast, $B_2$, which is targeted by $M_2(t)$, will likely depend on either itself, $B_3$ or possibly $B_4$ for its defence.}
\label{fig1}
\end{figure}

The context for this study is that of a team of AFVs being targeted by missile-equipped insurgents, who have fired several missiles towards the team. The  team has threat detection and defeat capability, such as APS. The purpose here is to quantify the team's survivability in the presence of this threat. Figure \ref{fig1} provides an illustration of the combat scene for a particular scenario. The AFV team consists of four vehicles, labelled $B_1, B_2, B_3$ and $B_4$. This figure illustrates the two missile case, which are denoted $M_1(t)$ and $M_2(t)$ respectively, where it is assumed that the team is stationary during the engagement. Missile 1 is fired toward $B_1$ while missile 2 is directed toward $B_2$, which are assumed to travel in a straight path to their targets. It is also assumed that the missiles are fired independently, and that the teams detect and disrupt the missiles independently. The concept of collaboration is understood to be implemented through automatic radar detection and tracking information being shared within the team through secure communication or networked information sharing. As will become apparent in the examples to be studied, it will be assumed that vehicles with detection capability will schedule vehicles with countermeasures once a threat is determined to be within the latter's strike range. This process was applied in the analysis presented in \cite{weinberg21}.

Introduce a continuous time stochastic process $\{Z(t), t\geq 0\}$ that takes values in the set $\{0, 1, 2\}$ where the number of missiles defeated at time $t$ is given by $Z(t)$.
Hence this is a jump process, beginning in state 0, and then spending sojourn times in states 0 and 1 before being absorbed into state 2, under the assumption that this process evolves continuously. In reality the process may never enter states 1 or 2; this depends on whether the AFV team is able to defeat missiles before they impact their targets. A simplifying assumption adopted is that the two missiles cannot be defeated at exactly the same time. Otherwise it will be possible to jump from state 0 to 2 directly. The process also cannot revert back to previous states by the assumption that there are only two missile threats. 

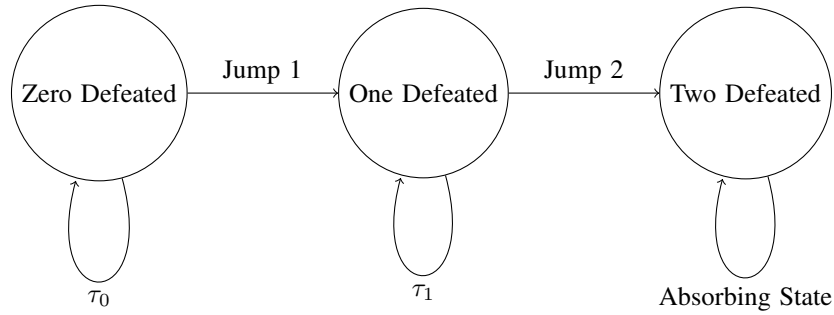
\begin{figure}[h]
\centering
\begin{tikzpicture}
[node distance=2cm]
        \node[state]             (s) {Zero Defeated};
        \node[state, right=of s] (r) {One Defeated};
          \node[state, right=of r] (t) {Two Defeated};

\path [->, draw] (s) -- node [text width=2.5cm,midway,above,align=center] {Jump 1} (r);
\path [->, draw] (r) -- node [text width=2.5cm,midway,above,align=center] {Jump 2} (t);
       
\draw[every loop]
(s) edge[loop below] node {$\tau_0$} (s)
(r) edge[loop below] node {$\tau_1$} (r)
(t) edge[loop below] node {Absorbing State} (t);
    \end{tikzpicture}
\caption{Transition diagram for the jump stochastic process $Z(t)$ in the case of two missiles. The process begins in state 0, remains there for $\tau_0$ and then jumps to state 1. It remains in the latter for a sojourn time $\tau_1$ and then finally ends up in its absorbing state, at time $\tau_0 + \tau_1$, of state 2.}
\label{fig2}
\end{figure}

Figure \ref{fig2} illustrates the jump states of the process. The process begins in state 0 where it remains for a sojourn time $\tau_0$, after which it then jumps to state 1 and remains there for $\tau_1$. Thereafter, it is absorbed into state 2, where it remains. The analysis presented in this paper quantifies the survivability of the team by considering how long it takes for the process $Z(t)$ to reach state 2, relative to the time it takes at least one missile to reach the team. This can be done by relating the sojourn time in states 0 and 1 to the jumps of the process $Z(t)$ through conditional probability. Initially generic functions of time will be used for the probability of detection of the missile, as well as for the probability of disrupting it given it has been detected. In the specific examples of Section \ref{sec:5} particular choices for these will be introduced to produce performance prediction results.

The key idea utilised in this analysis is that the time the process takes to reach state 2 will determine whether the AFV team has defeated both missile threats. Hence with $\tau_0$ and $\tau_1$ as defined previously  the random variable $\tau_0 + \tau_1$ is the time it takes for the process $Z(t)$ to reach state 2, which means that both missiles are defeated. If this random variable is smaller than the time it takes for the faster missile to reach its target, under the assumption that the missile is not defeated, then it follows that the team has succeeded in neutralising the threat. Thus the distribution of the sum of sojourn times can be used to quantify the performance of the C-APS at any given time instant.
The following section illustrates this in the simple case of a single missile threat.

\section{Single Missile Case}
\label{sec:3}
Suppose that the AFV team is facing a single missile threat, which if not intercepted, will reach its intended target at time $T$ seconds. Suppose that $D(t)$ is the probability that the missile is detected at time $t$. Similarly, let $C(t)$ be the probability that the missile is countered or  defeated at time $t$, given it has been detected. Throughout the analysis it will be assumed that detections and countering are independent over different time epochs, in order to avoid cascading conditional probabilities. The process  $Z(t)$ takes the value 1 if the missile is defeated at time $t$, and zero otherwise. Suppose that $\tau_0$ is the time the process spends in state 0 before it exits to state 1. The idea proposed in this analysis is that if the time spent in state 0 is smaller than the time it takes the missile to arrive at its intended target, under the assumption it is not defeated, then the team will have defeated the missile and survived the attack. Hence one can examine the complementary distribution function of $\tau_0$ to quantify the AFV's C-APS performance.

Observe that if $\tau_0 > t$ then the process is still in state 0 at time $t$. Similarly, if the process is still in state 0 at time $t$ then its sojourn time in this state is at least $t$.
Hence the event $\{\tau_0 > t\}$ is equivalent to $\{Z(t) = 0 | Z(0) =0\}$, and thus
\begin{equation}
\prob(\tau_0 > t) = \prob(Z(t) = 0 | Z(0) = 0).
\label{auxeq1}
\end{equation}
Given the process begins in state 0, it will be still in that state at time $t$ under two conditions. Either the missile is not detected at time $t$ or if it is detected then it is not defeated. Hence
\begin{equation}
\prob(Z(t) = 0 | Z(0) = 0) = (1-D(t)) + D(t)(1-C(t)) = 1-D(t)C(t).
\label{auxeq2}
\end{equation}
Equivalently, the product $D(t)C(t)$ is the probability that the missile is detected and countered at time $t$, so its complement is the probability that the process $Z(t)$ remains in state 0 at time $t$.

An application of \eqref{auxeq2} to \eqref{auxeq1} establishes that
\begin{equation}
\prob(\tau_0 > t)  = 1-D(t)C(t).
\label{auxeq3}
\end{equation}
Thus \eqref{auxeq3} is the complementary distribution function of the time spend in state 0 before the process exits to state 1. If $\tau_0 < T$ then the team has defeated the missile before it has intercepted its target. Hence the complementary distribution function \eqref{auxeq3} can be used to quantify the performance of the  team in terms of defeating the missile threat.

By taking the complement of \eqref{auxeq2} observe that
\begin{equation}
\prob(Z(t) = 1 | Z(0)  = 0) = D(t)C(t), \label{neq1}
\end{equation}
since the process $Z(t)$ only takes values in the set $\{0, 1\}$. Therefore \eqref{neq1} can also be used to quantify the AFV team's survivability at any given time $t$, since this is the probability that the AFV team has defeated the threat by time $t$.

 It will become apparent, as the analysis proceeds, that expressions \eqref{auxeq3}  and \eqref{neq1} provides a significantly simpler solution to the single missile case examined in \cite{weinberg21}. In the next section the extension of this approach, to the scenario of two missiles threatening the AFV team, will be examined.

\section{Two Missile Case}
\label{sec:4}
The extension of the approach in the previous section to two missiles introduces a degree of complexity, but nonetheless a useful solution can be constructed.
The stochastic process $Z(t)$ takes values in the set $\{0, 1, 2\}$ with sojourn times $\tau_0$ and $\tau_1$ for the time it spends in states 0 and 1 respectively.
Suppose that the two missile takes $T_1$ and $T_2$ seconds respectively to reach their targets if they are not disrupted. 
Then if $\tau_0 + \tau_1 < \min\{T_1, T_2\}$ both missiles are defeated; the complement of this event implies at least one missile has reached the team. 
Hence, as in the single missile case, the complementary distribution function of the sum of sojourn times can be used to quantify the likelihood that the team survives the attack.

Note that if $\tau_0 + \tau_1 < t$ then one can conclude that the process $Z(t)$ is in its absorbing state 2 at time $t$. Similarly, if $Z(t) = 2$ then the sum of sojourn times cannot exceed $t$. Hence it follows that 
\begin{equation}
\prob(Z(t) = 2 | Z(0) = 0) = \prob(\tau_0 + \tau_1 < t). \label{neq2}
\end{equation}
Therefore one can utilise \eqref{neq2} to predict the likelihood that the AFV team has defeated the two missiles by time $t$. 

The complementary distribution function of the sum of sojourn times is now derived.
Observe that by conditioning on $\tau_0$, and for fixed $t > 0$,
\begin{equation}
\prob(\tau_0 + \tau_1 > t) = \int_0^\infty \prob(\tau_0 + \tau_1 > t | \tau_0 = s) f_{\tau_0}(s) ds \label{condexp1}
\end{equation}
where $f_{\tau_0}$ is the density of $\tau_0$. Note that the probability in the intergrand of \eqref{condexp1} is unity in the case where $s > t$. Thus \eqref{condexp1} is equivalent to 
\begin{eqnarray}
\prob(\tau_0 + \tau_1 > t) &=& \int_0^t \prob(\tau_1 > t - \tau_0 | \tau_0 = s) f_{\tau_0}(s) ds 
 + \int_t^\infty f_{\tau_0}(s) ds \nonumber\\
&=& \int_0^t \prob(\tau_1 > t - s | \tau_0 = s) f_{\tau_0}(s) ds + \prob(\tau_0 > t). \label{condexp2}
\end{eqnarray}
In view of \eqref{condexp2} it is necessary to determine the complementary distribution function of $\tau_0$, its density, and the distribution of $\tau_1$ conditioned on $\tau_0$.
These can be determined by relating the sojourn times $\tau_0$ and $\tau_1$ to jump probabilities of the process $Z(\cdot)$ and the probabilities of detection and defeat of the incoming missiles. Towards this objective, let $D_j(t)$ be the probability that missile $j$ is detected at time $t$, and $C_j(t)$ be the probability that missile $j$ is countered at time $t$, given it has been detected. It will be assumed that these functions are both differentiable and integrable for all $t \geq 0$.

As in the single missile case, one can determine the validity of \eqref{auxeq1} also for the context of two or more missiles; the transition probability on the right hand side of \eqref{auxeq1} will change though.
If the process is in state 0 at time $t$ then neither missile has been detected and defeated. Hence it follows that
\begin{equation}
\prob(\tau_0 > t) = [1-D_1(t)C_1(t)][1-D_2(t)C_2(t)]. \label{tau0prob}
\end{equation}

There is merit in noting that the probability in \eqref{tau0prob} is exactly that of the process $Z(t)$ being in state 1, given it began in state 0. However the complement of \eqref{tau0prob} is
\begin{equation}
\prob(\tau_0 \leq t) = \prob(Z(t) = 1 \mbox{ or } 2 | Z(0) = 0)\label{neq3}.
\end{equation}
Hence \eqref{neq3} can be used to quantify whether the AFV team has defeated one of the two missile by time $t$.

The density of $\tau_0$ can be constructed from the above, since $f_{\tau_0}(t) = - \frac{d}{dt} \prob(\tau_0 > t)$. However, this will not be required for the analysis to follow.

Given that $\tau_0 = s$, the event that $\tau_1$ exceeds $t$ will imply that the original process $Z(\cdot)$ will be in state 1 at time $t+s$. This can also be interpreted as the event that $Z(t+s) = 1$ conditioned on  $Z(s) = 1$ and $\tau_0 = s$.  By a similary argument the converse can be demonstrated to be true. Hence it follows that
\begin{equation}
\prob(\tau_1 > t | \tau_0 = s) = \prob(Z(t+s) = 1 | Z(s) = 1, \tau_0 = s). \label{tau1}
\end{equation}
The evolution of the process $Z(\cdot)$, after it has exited state 0, is only dependent on the time it spend in that state, which is captured by its dependency on $s$ in \eqref{tau1}. 
Given $Z(s) = 1$, the event $Z(t+s) = 1$ implies either the second missile has not been detected in the time interval $[s, t+s]$ or if it has been detected, then it has not been defeated at time $t+s$. The condition that $\tau_0 = s$ implies that at time $s$ one of the missiles has been defeated; it is impossible to know {\em a priori} which one it is. Thus with an application of the law of total probability conditioned on events,
\begin{eqnarray}
\prob(Z(t+s) = 1 \!\!\!\!\!\!\!&|&\!\!\!\!\!\!\! Z(s) = 1, \tau_0  = s) \nonumber\\
&=& \prob(Z(t+s) = 1 | Z(s) = 1, \tau_0 = s, \mbox{missile 1 defeated at time } s) \times\nonumber\\
&&\prob(\mbox{missile 1 defeated at time }s | Z(s) = 1, \tau_0 = s)\nonumber\\
&& + \prob(Z(t+s) = 1 | Z(s) = 1, \tau_0 = s, \mbox{missile 2 defeated at time } s)\times\nonumber\\
&&\prob(\mbox{missile 2 defeated at time }s | Z(s) = 1, \tau_0 = s). \label{revcond1}
\end{eqnarray}
Conditioned on $Z(s) = 1, \tau_0 = s$ and missile 1 being defeated at time $s$, the probability that $Z(t+s) = 1$ is the probability that at time $t+s$ the second missile has not been defeated. Hence 
\begin{equation}
\prob(Z(t+s) = 1 | Z(s) = 1, \tau_0 = s, \mbox{missile 1 defeated at time } s)  = 1- D_2(s+t)C_2(s+t), \label{revcond2}
\end{equation}
and similarly
\begin{equation}
\prob(Z(t+s) = 1 | Z(s) = 1, \tau_0 = s, \mbox{missile 2 defeated at time } s)  = 1- D_1(s+t)C_1(s+t). \label{revcond3}
\end{equation}
Next, the probability that missile 1 is defeated at time $s$ given $Z(s) = 1$ and $\tau_0 = s$ is equivalent to the probability that
missile 1 is defeated at time $s$ given either missile 1 or missile 2 are defeated at time $s$. In this context, missile 1 defeated is contingent on missile 2 surviving and so must be included in the calculation.
Due to the fact that $\tau_0 = s$, the event missile 1 is defeated and the event missile 2 is defeated are mutually exclusive, and so
\begin{eqnarray}
\prob(\mbox{missile 1 defeated at time }s \!\!\!\!\!\!\!&|&\!\!\!\!\!\!\! Z(s) = 1, \tau_0 = s)\nonumber\\ 
&=&  \frac{\prob(\mbox{missile 1 defeated at time }s)}{ \prob(\mbox{missile 1 defeated at time }s) + \prob(\mbox{missile 2 defeated at time }s)}\nonumber\\
&=& \frac{D_1(s)C_1(s)[1-D_2(s)C_2(s)]}{D_1(s)C_1(s)[1-D_2(s)C_2(s)]+D_2(s)C_2(s)[1-D_1(s)C_1(s)]}.\nonumber \\
 \label{revcond4}
\end{eqnarray}
By taking the complement of \eqref{revcond4} it can be seen that
\begin{eqnarray}
\prob(\mbox{missile 2 defeated at time }s \!\!\!\!\!&|&\!\!\!\!\! Z(s) = 1, \tau_0 = s) \nonumber\\
&&= \frac{D_2(s)C_2(s)[1-D_1(s)C_1(s)]}{D_1(s)C_1(s)[1-D_2(s)C_2(s)]+D_2(s)C_2(s)[1-D_1(s)C_1(s)]}.\nonumber\\
 \label{revcond5}
\end{eqnarray}
Finally, by applying \eqref{revcond2} - \eqref{revcond5} to \eqref{revcond1} it follows that
\begin{eqnarray}
\prob(Z(t+s) = 1 | Z(s) =1, \tau_0  = s)  &=& \Bigg\{D_1(s)C_1(s)[1-D_2(s)C_2(s)][1-D_2(s+t)C_2(s+t)]  \nonumber\\
&& + \, D_2(s)C_2(s)[1-D_1(s)C_1(s)][1-D_1(s+t)C_1(s+t)]\Bigg\}\nonumber\\
&\div& \nonumber\\
 && \Bigg\{D_1(s)C_1(s)[1-D_2(s)C_2(s)]+D_2(s)C_2(s)[1-D_1(s)C_1(s)]\Bigg\}.\nonumber\\
\label{revcond6}
\end{eqnarray}

This can now be applied to \eqref{tau1}, and when the result is substituted into \eqref{condexp2} one produces
\begin{eqnarray}
\prob(\tau_0  + \tau_1 > t) &=& [1-D_2(t)C_2(t)]\times \nonumber\\
&&\int_0^t\frac{ D_1(s)C_1(s) [1-D_2(s)C_2(s)]}{D_1(s)C_1(s)[1-D_2(s)C_2(s)]+D_2(s)C_2(s)[1-D_1(s)C_1(s)]}f_{\tau_0}(s) ds \nonumber\\
&& + [1-D_1(t)C_1(t)]\times\nonumber\\
&&\int_0^t \frac{D_2(s) C_2(s) [1-D_1(s)C_1(s)]}{D_1(s)C_1(s)[1-D_2(s)C_2(s)]+D_2(s)C_2(s)[1-D_1(s)C_1(s)]}f_{\tau_0}(s) ds \nonumber\\
&& + [1-D_1(t)C_1(t)][1-D_2(t)C_2(t)],\nonumber\\
\label{finaltaudis}
\end{eqnarray}
where \eqref{tau0prob} has also been utilised.

Hence \eqref{finaltaudis}  can be used to quantify the probability that the team defeats the two missiles. This can be done by an application of it to \eqref{neq2}.

In order to evaluate the integrals in \eqref{finaltaudis} two approaches can be applied. The first, and most direct technique, is to apply the density of $\tau_0$ to this integral, followed by numerical integration. However, this will require the derivative of the functions $D_j(t)$ and $C_j(t)$ to be determined. Numerical differentiation will therefore be necessary.

A second method is to note that the integrals in \eqref{finaltaudis} can be written
\begin{equation}
\int_0^t \frac{ D_j(s)C_j(s)[1-D_k(s)C_k(s)]}{q(s)} f_{\tau_0}(s) ds = \int_0^\infty \I[s \leq t] \frac{D_j(s)C_j(s) [1-D_k(s)C_k(s)]}{q(s)}f_{\tau_0}(s) ds \label{temp1}
\end{equation}
where $j, k \in \{1, 2\}, j \not = k, $ and $ \I[s \leq t] $ takes the value unity if $s \leq t$ and is zero otherwise, and for brevity
\begin{equation}
q(s) : = D_1(s)C_1(s)[1-D_2(s)C_2(s)]+D_2(s)C_2(s)[1-D_1(s)C_1(s)]. \label{qfundef}
\end{equation}
Then the integral \eqref{temp1} is the mean of 
\begin{equation}
g(s) := \I[s \leq t] \frac{D_j(s)C_j(s)[1-D_k(s)C_k(s)]}{q(s)}
\label{gfundef}
\end{equation}
with respect to $\tau_0$. Hence it follows that
\begin{equation}
\int_0^t \frac{D_j(s)C_j(s) [1-D_k(s)C_k(s)]}{q(s)}f_{\tau_0}(s) ds = \mean(g(\tau_0)). \label{temp2}
\end{equation}
Consequently one can apply a Monte Carlo scheme to approximate \eqref{temp2} by realising values of the random variable $\tau_0$ and applying
\begin{equation}
\mean(g(\tau_0)) \approx \frac{1}{G}\sum_{j=1}^G g(\tau_0^{(j)}), \label{temp3}
\end{equation}
where $G$ is the number of Monte Carlo simulation runs and  $\tau_0^{(j)}$ are successive realisations of $\tau_0$. Since the distribution function of $\tau_0$ is the complement of \eqref{tau0prob} these realisations can be obtained numerically by simply generating uniform random numbers in the unit interval and finding the solution to the equation $\prob(\tau_0 > t) = r_j$, where $r_j$ is a uniformly distributed random variable realisation in the unit interval. 

The solution to the two missile case has increased in complexity from the one missile scenario, and now requires Monte Carlo estimation to facilitate the evaluation of integrals. 
In the next section some examples of predicted performance are provided. Since the AFV defence will be provided by a HEL DEW it is necessary to produce a suitable expression for the probability of disruption of a given target.

\section{Examples of Performance}
\label{sec:5}
The purpose of this section is to provide some tangible examples of performance of AFV survivability in the presence of missile threats. Detection of targets and their tracking will be adopted as in the approach of \cite{weinberg21}, while active defence of targets will be provided through a HEL DEW. The latter necessitates the derivation of a novel expression for the probability of DEW effect on a target, which also includes a weapon dwell time.

\subsection{Directed Energy High Energy Laser Defence}
DEWs are an emerging disruptive technology that have the potential to provide effective defence against small targets, such as missiles and unmanned aerial vehicles \cite{nielsen}. The two main categories of DEWs are HPRF, which can provide a soft-kill option, and HELs, which can be used to deliver a hard-kill capability \cite{deveci}. The applications in \cite{weinberg21} focused on HPRF DEWs, which have been shown to have the potential to disrupt a missile's guidance and control, and especially its motor \cite{radasky, clarke, radasky2}. 
HEL DEWs disrupt a missile by causing a thermal effect which can then cause a missile to become unstable in flight and consequently self-destruct \cite{pluent17}. When targeting a UAV the same effect can be produced. The application of HEL DEWs for short range defence has been examined in \cite{lavan},  while its integration with land vehicles has been investigated in \cite{hafften}. HEL DEWs can also provide a ship with defence capability, which has been documented in  \cite{ang}.  There are several commerically available HEL DEWs being prototyped by industry. A series of significant exemplars have been developed by the German Rheinmetall Group \cite{rheinmetall1}. These include their 5 kW HEL, installed on a GTK Boxer wheeled tank, a 1 kW HEL on an M113 tracked vehicle and a 20 kW HEL on an armoured Tatra truck \cite{rheinmetall2}. It is of interest to note that Rheinmetall reported that the 5 kW HEL was used to neutralise an extra-heavy machine gun mounted on the bed of a travelling small truck, without injuring the marksman. Sensors mounted in the vehicle provided data to demonstrate that the HEL DEW could be used for precision strikes with minimal collateral damage. In addition to this, Rheinmetall reports that its Air Defence HEL 30 kW effector was able to neutralise five 82 mm mortar shells, at a distance of 1 km, in roughly four seconds. This clearly has demonstrated the utility of HEL effectors for defence of combat vehicles.

The way in which a DEWs effectiveness for target disruption can be measured is through its power density, in Watts per metre square, it delivers on the target. The following will focus on that for HEL DEWs but it is worth observing that the discussion can be adapted quite easily to include HPRF DEWs. However, it is important to note that the propagation of a laser beam is subjected to a number of distortion and environmental attenuating factors, such as diffraction, jitter, atmospheric turbulence, scattering, scintillation and thermal blooming \cite{cook, gebhardt}. 

For a HEL DEW, the peak power density on a target at time $t$ can be expressed as
\begin{eqnarray}
I(t) &=& \left( \frac{6.28 P_0}{{\cal M}^4 + 2.9 \left( \frac{R_0}{r_0}\right)^2 + \left( \frac{\pi R_0 \theta_J}{\lambda}\right)^2}\right) 
\left( \frac{R_0}{\lambda R(t)}\right)^2 e^{-\gamma R(t)}, \label{pdf1}
\end{eqnarray}
where $P_0$ is the output power of the DEW (in Watts), $\lambda$ is the wavelength (in metres), 
$R(t)$ is the range to the target (in metres), $R_0$ is the initial laser spot size, ${\cal M}^2 $ is the intrinsive laser beam quality, $r_0$ is the transverse coherence length associated with turbulence and $\theta_J$ is the mechanical jitter angle of the beam. The coefficient $\gamma$ represents atmospheric extinction due to both absorbtion and scattering. The power density \eqref{pdf1} has been adopted from \cite{sprangle}, and suitable parameters for it have been taken from \cite{phillip}. The coefficient $r_0$ is given by
\begin{equation}
r_0 = 0.184 \left( \frac{\lambda^2}{c_n^2 R(t)}\right)^{\frac{3}{5}}, \label{pdf1a}
\end{equation}
where $c_n^2$ characterises the strength of turbulence \cite{hafizi}.

Based upon the example provided in Tables 1 and 2 of \cite{phillip} the following HEL DEW parameters have been adopted. The wavelength $\lambda = 1.045 \, \mu$m,  $\theta_J = 1$, ${\cal M}^2 = 4$, $c_n^2 = 10^{-15}$ and $\gamma = 2\times 10^{-3} + 1.2\times 10^{-1}$ km$^{-1}$. In contrast to \cite{phillip} the laser power is selected to be 30 kW to match Rheinmetall's HEL characteristics. Also the initial laser spot size has been selected to be $R_0 = 0.1$ m since the target will be small. By applying these parameters to \eqref{pdf1} and \eqref{pdf1a} it can be shown that
\begin{equation}
I(t) = \frac{172.5179 R^{-2}(t) e^{-1.22\times 10^{-4} R(t)}}{ \left(16\times 10^{-13}\right) + \left(9.0379\times 10^{-3}\right) + \left(3.2\times 10^{-16} R^{\frac{6}{5}}(t)\right)}. \label{pdf1b}
\end{equation}
Since the application considered in this paper is a team of AFVs being atttacked by insurgents, it is reasonable to assume that the engagement will be close combat, so that the target fired will be within a few kilometres of the team. This implies that the exponential term contribution to the numerator in \eqref{pdf1b} can be considered negligible, as well as the term in the denominator involving $R(t)$. Thus \eqref{pdf1b} can be approximated by
\begin{equation}
I(t) = \frac{19.0883 \times 10^3}{R^2(t)}. \label{pdf1c}
\end{equation}
It is interesting to observe that the expression \eqref{pdf1c} is somewhat similar to the corresponding expression for the HPRF DEW power density function \cite{weinberg21}.

Both HEL, as well as HPRF, DEWs require a dwell time on its intended target. This is more certainly the case with a HEL DEW since it will take time to cause the surface temperature of the target to rise, in order to create a thermal disruption. The method in which dwell time is modelled in this study is to assume it has a distribution that is a function of the time it is likely for the DEW to have power concentrated on a target. Hence, supposing that the interest is in measuring the impact of a DEW at time $t$ where it is assumed that at this time the target has been detected, let $\tau$ be the dwell time. Then this random variable takes values in the interval $[0, t]$, since either the DEW's target is first radiated at time $t$ or it has been radiated for some time within this time interval. This approach avoids having to apply sequential conditional probability arguments for the probability of detection, and target illumination by the DEW, over time prior to $t$.

The power density over the radiating window of $[t-\tau, t]$ is given by
\begin{equation}
 \int_{t-\tau}^t I(s) \frac{(1-r_f)}{c_{m}\rho_m h_m}ds, \label{pdf2}
\end{equation}
where $r_f$ is the surface reflectivity of the material, $c_m$ is its heat capacity, $\rho_m$ is its density and $h_m$ is the shell thickness. This expression has been adopted from \cite{yun20}, and following the latter's selection of parameters for the missile in their study, it is assumed that $r_f = 0.3$, $c_m = 900$  Joules per Kelvin, $\rho_m = 2.7 \times 10^3$ kg per m$^3$ and $h_m = 0.01$ m. Based upon these selections it follows that \eqref{pdf2} becomes
\begin{equation}
 2.88 \times 10^{-5} \times \int_{t-\tau}^t I(s) ds.\label{pdf2a}
\end{equation}
The target will have an effective area $\sigma$ where \eqref{pdf2a} will have the greatest impact. In terms of radio frequency, this corresponds to the radar cross section of the target, and as such is modelled through a statistical distribution. Hence the normalised power density, over the observational window, is given by the random variable
\begin{equation}
I_T(t) =  2.88 \times 10^{-5} \times \sigma \times  \int_{t-\tau}^t I(s) ds. \label{pdf3}
\end{equation}
When $I_T$ exceeds a threshold level then the target of the DEW is disrupted. Hence the probability of disrupting the target, given it was detected at time $t$ is 
\begin{equation}
q(t) = \prob\left( I_T(t) > U\right), 
\label{pdf4}
\end{equation}
where $U$ is a fixed threshold, in units of Watts per square metre per Kelvin. The latter threshold can be determined by knowledge of the temperature required to create a thermal effect with a HEL, or through knowledge of coupling power requirements to disrupt electronics for HPRF DEWs.

It is reasonable to assume that the DEW dwell time and the target effective area are statistically independent. Since $\tau$ is modelling dwell time, it is assumed that it has an exponential distribution, since the latter is a model for waiting times between events in a Poisson process \cite{ross}, which is a model for rare events. Since this distribution is limited to the time interval $[0, t]$ it is assumed that $\tau$ has density
\begin{equation}
f_{\tau}(s) = \frac{\mu_{\tau} e^{-\mu_{\tau}s}}{ 1 - e^{-{\mu_{\tau}}t}}, \label{pdf5}
\end{equation}
for $0 \leq s \leq t$, where the reciprocal of $\mu_{\tau}$ is the expected dwell time. 
It will also be assumed that the effective area of the target has an exponential distribution. Specifically, $\sigma$ has density
\begin{equation}
f_{\sigma}(s) = \mu_{\sigma} e^{-{\mu_\sigma}s}, \label{pdf6}
\end{equation}
for $s \geq 0$, where the reciprocal of $\mu_{\sigma}$ is the mean effective area. Such a model, in radar signal processing, corresponds to assuming a Swerling I target model, and is adopted for simplicity in the current analysis.

Observe that by conditioning on $\tau$, the probability \eqref{pdf4} can be expressed
\begin{equation}
q(t) = \int_0^t \prob( I_T(t) > U | \tau = s) f_{\tau}(s) ds.
\label{pdf7}
\end{equation}
By applying \eqref{pdf5} and \eqref{pdf6} to \eqref{pdf7} it can be shown that the latter reduces to 
\begin{equation}
q(t) = \frac{\mu_{\tau}}{1-e^{-{\mu_{\tau}} t}} \int_0^t e^{-{\mu_\tau}s}   
e^{ -1.8194\mu_{\sigma}U  \left[ \int_{t-s}^t R^{-2}(x) dx\right]^{-1}} ds, \label{pdf8}
\end{equation}
where \eqref{pdf1c} has also be utilised.

Consequently \eqref{pdf8} provides an expression for the probability of disruption of a target by a HEL DEW. This expression accounts for the power delivered on the target (which is essentially the kill probability) and also for the likelihood that the DEW concentrates its power on a target's vulnerable area (the probability of hit).

\subsection{Radar Detection Assumptions}
Given the problem of interest is line of sight detection in a land environment, it is reasonable to adopt the assumptions employed in \cite{weinberg21}.
Hence it is assumed that the radar bases its detection on a series of $M$ pulses, that are non-coherently integrated in response. Due to these conditions, the probability of detection of the target is given by the Marcum Q-function defined by 
\begin{equation}
p = Q_M(\sqrt{2M \zeta}, \sqrt{2\nu}), 
\label{mqf1}
\end{equation}
where 
\begin{equation}
Q_M(a, b) = \frac{1}{a^{M-1}} \int_b^\infty  x^M \exp\left(-\frac{(x^2 + a^2)}{2}\right) I_{M-1}(ax) dx
\end{equation}
and $I_M$ is the modified Bessel function of the first kind of order $M$, 
$\zeta$ is the signal to clutter ratio (SCR) and $\nu$ is the detection threshold, given by the solution to the equation
\begin{equation}
\pfa = Q_M(0, \sqrt{2\nu}) \label{mqf2}
\end{equation}
where $\pfa$ is the probability of false alarm. The SCR can be specified as a function of radar characteristics through
\begin{equation}
\zeta = \frac{P_R G_R^2 \lambda_R \mean(\sigma_{R})}{(4\pi)^3 R^4  2\theta}, \label{scr}
\end{equation}
where $P_R$ is the power radiated by the radar (in Watts), $G_R$ is the radar antenna's gain (in dBi), $R$ is the distance to the target (in metres), $\lambda_R$ is the wavelength (in metres) of the radar signal, $\mean(\sigma_R)$ is the mean of the RCS $\sigma_R$ of the target (in square metres)  and 
$\theta$ is the variance of the compound Gaussian model in the underlying assumption of Rayleigh amplitude statistics.  
The issue with applying the Marcum Q-function to quantify detection performance is that it has been developed to model lower resolution radars, and consequently it does not necessarily reflect contemporary radar performance. Therefore, for the current study, radar operational parameters have been selected so that detection of a small target at a 1 km range is likely, to reflect performance cited by Rheinmetall. Hence it is assumed that $P_R = 10^8$ Watts with a gain of $G_R = 100$ dBi and a wavelength of $\lambda_R = 0.03$, where the latter corresponds to a frequency of 10 GHz.  The average radar target cross section has been chosen to be unity, so that $\mean(\sigma_{R}) = 1$. It will also be assumed that $M=32$ with $\pfa = 10^{-4}$.

Selection of the parameter $\theta$ will be based upon producing an effective detection range of roughly 1 km, for the target of the missile. For the examples to be considered, it was found that $\theta = 10^{-5}$ provided this expected effective detection range with the Marcum $Q$-Function.

The next subsection discusses the probability of detection and disruption for the team in terms of \eqref{mqf1} and \eqref{pdf8}.

\subsection{Detection and Disruption Probabilities}
It is now possible to specify the functions $D_j(t)$ and $C_j(t)$, introduced in previous sections to model the probability of detection and disruption of missile $j$ at time $t$.
Suppose that the team contains $L$ AFVs, who are members of the set ${\cal B} =\{B_1, B_2, \ldots, B_L\}$.  Suppose that a series of missiles are fired toward the team. Define a set ${\cal D}$ to be
\begin{equation}
{\cal D} = \{B_j \in {\cal B}: B_j \mbox{ has target detection and tracking capability}\}. \label{pdf9}
\end{equation}
Similarly, define a time-dependent set 
\begin{equation}
{\cal S}_i(t) = \{B_j \in {\cal B}: B_j \mbox{ is a potential disruptor of missile $i$ at time $t$}\}. \label{pdf10}
\end{equation}

Next define $p_{i, j}(t)$ to be the probability that missile $i$ is detected by $B_j \in {\cal D}$ at time $t$. Also define $q_{i, j}(t)$ to be the probability that missile $i$ is disrupted by
 $B_j \in {\cal S}_i(t)$ at time $t$, given it has been detected.

Then these individual probabilities can be used in the formulation of team detection and defeat as in the discrete time case in \cite{weinberg21}.
For the case of detection, 
\begin{eqnarray}
D_i(t) &=& \prob(\mbox{Missile $i$ detected by team at time $t$}) \nonumber\\
&=& 1 - \prod_{ \{j: B_j \in {\cal D}\}}[1-p_{i, j}(t)], \label{pdf11}
\end{eqnarray}
while for disruption
\begin{eqnarray}
C_i(t) &=& \prob(\mbox{At least one } B_j \in {\cal S}_i(t) \mbox{ defeats the missile at time } t)\nonumber\\
&=& 1 - \prod_{ \{j: B_j \in {\cal S}_i(t)\}}[1-q_{i, j}(t)]. \label{pdf12}
\end{eqnarray}

Expressions \eqref{pdf8} and \eqref{mqf1} can now be applied to \eqref{pdf12} and \eqref{pdf11} respectively. 

\subsection{Single Missile Attack Performance Prediction}
It is assumed that the team consists of four AFVs, and that the missile is fired toward $B_1$, as illustrated in Figure \ref{fig1}. $B_2$ is located on the $x$-axis as shown, at a distance of 10 m from $B_1$. Vehicles $B_3$ and $B_4$ are located at angles of $60^\circ$ to the $y$-axis, with $B_3$ 50 m from $B_1$ and $B_4$ at a distance of 30 m. It is assumed that the missile $M_1(t)$ travels to $B_1$ at a constant speed $\nu$, and will strike it, if not intercepted, at time $T$. Hence the distance of the missile from $B_1$ at time $t$ is
\begin{equation}
R_1(t) = \nu(T-t), \label{R1}
\end{equation}
for $0 \leq t \leq T$.
The other associated distances can be produced by an application of the cosine rule, noting that it is assumed that during the engagement the AVFs remain stationary. Hence if $R_j(t)$ is the distance between vehicle $j$ and the missile at time $t$ then
\begin{eqnarray}
R_2(t) &=& \sqrt{\nu^2(T-t)^2 + 100} \label{R2}\\
R_3(t) &=& \sqrt{(\nu(T-t) - 25)^2 + 1875} \label{R3}\\
R_4(t) &=& \sqrt{ (\nu(T-t) - 15)^2 + 675},\label{R4}
\end{eqnarray}
provided $0 \leq t \leq T$.
These expressions can now be applied to \eqref{pdf8} and  \eqref{mqf1}, with the results then utilised in \eqref{pdf11} and \eqref{pdf12}.

As in the approach in \cite{weinberg21}, the average target illumination parameter $\mu_\sigma$ can be related to the radar's average cross section $\mean(\sigma_R)$ by consideration of the surface area and normal to it. In particular, the average target cross section is given by $\frac{4\pi {\cal A}^2}{\lambda^2}$, where ${\cal A}$ is the target's surface area and $\lambda$ is the wavelength of the incident wave. If it is assumed that the radar cross section is unity, then the parameter $\mu_\sigma$ can be shown to be given by
\begin{equation}
\mu_\sigma = \frac{1.045^2}{9} \times 10^{-8}. \label{musig}
\end{equation}

For the HEL DEW the average dwell time on the target is assumed to be 3 seconds, so that $\mu_{\tau} = 3$. The missile is assumed to travel at a speed of $\nu = 30$ metres per second, and it will take $T = 120$ seconds to reach $B_1$ if not intercepted. Hence it is fired from a distance of 3.6 km.
Numerical experimentation demonstrated that with a choice of $\theta = 10^{-5}$ then a radar located on vehicle $B_1$ will detect the incoming missile with a probability of 0.8703 at a range of 900 m. Hence this choice has been selected for this example. 

\begin{figure}[h]
\centering
\includegraphics[width=13cm]{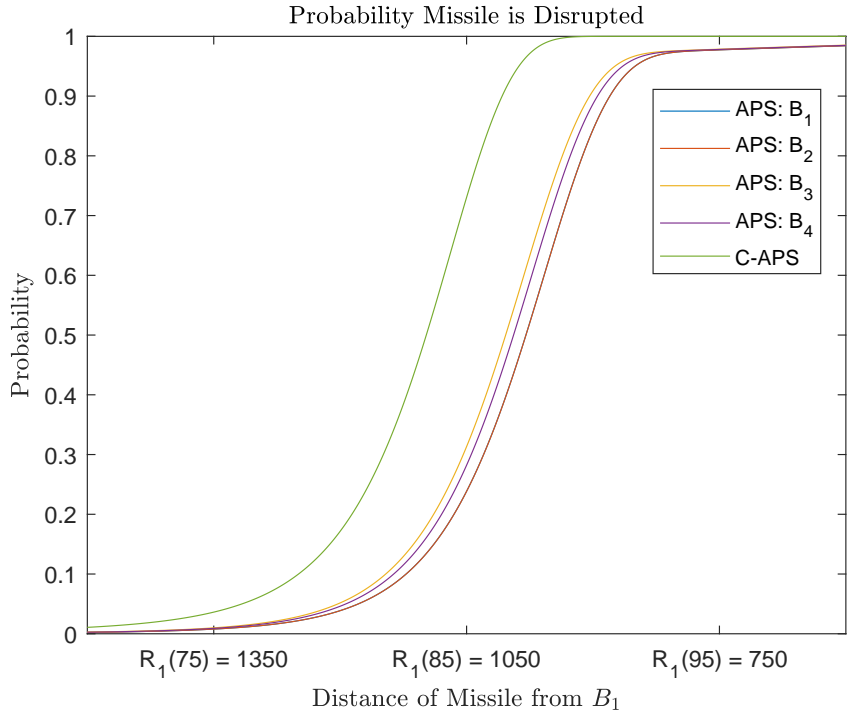}
\caption{Performance prediction of individual APS and C-APS for the case where the disruption threshold is $U = 10$ Watts per metre squared per Kelvin. The $x$-axis plots the distance of the missile to its target ($B_1$) as a function of time, namely $R_1(t)$ (units of metres). }
\label{fig1a}
\end{figure}

\begin{figure}[h]
\centering
\includegraphics[width=13cm]{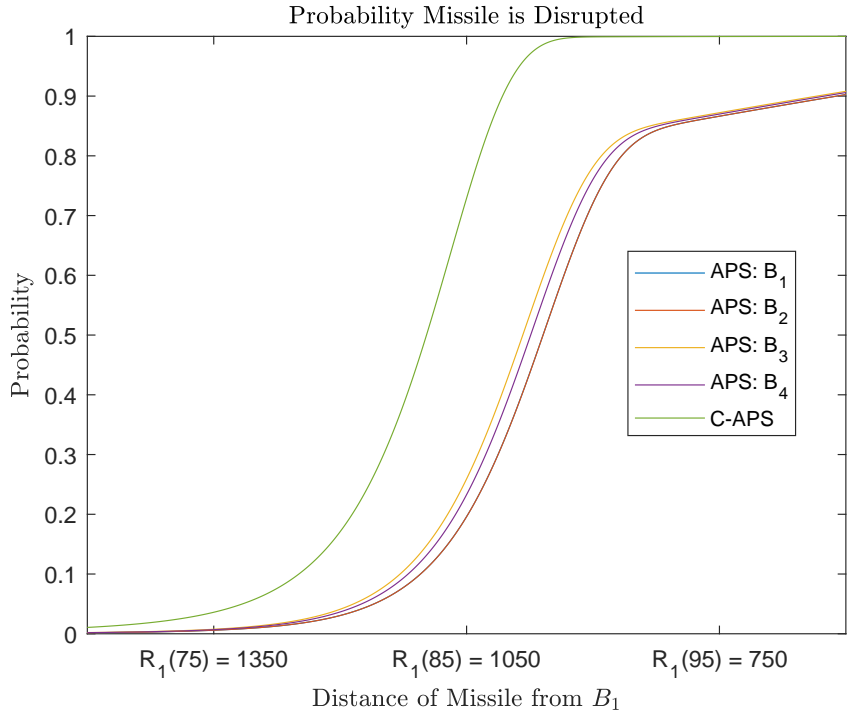}
\caption{Similar plot to Figure \ref{fig1a} except the HEL disruption threshold has been increased to $U = 100$. A point of interest is that the collaborative APS tends to perform much the same as for the example in Figure \ref{fig1a} but clearly the individual member's APS performance has declined.}
\label{fig1b}
\end{figure}

Figures \ref{fig1a} and \ref{fig1b} provide two examples of predicted performance. In these examples, the case where each individual vehicle is equipped with APS alone is examined, and then it is compared with the case where each vehicle is equipped with coordinated detection and disruption. These figures plot the distribution function of $\tau_0$, which is equivalent to the probability that the process $Z(t)$ is in state 1 at time $t$, given it started in state 0. Hence the figure can be used to assess the likelihood that the AFVs defeat the missile by a given time $t$.

Figure \ref{fig1a} is for the case where the disruption threshold is $U = 10$, while in Figure \ref{fig1b} this has been increased to $U  = 100$. 
The $x$-axis plots the distance the missile has to travel to reach $B_1$ at time $t$; hence it is plotting $R_1(t)$. 
Comparing these figures one observes that individual APS performance drops as the threshold increases, which is to be expected. However, for the full C-APS capability, the performance remains almost identical. When $U$ is increased significantly, there is an overall decline in predicted performance, which is to be expected (not shown for brevity).

In terms of the missile's distance from $B_1$, at time 75 it at 1350 m, at time 85 it is 1050 m and at time 95 it is 750 m from $B_1$. 
Hence Figure \ref{fig1a} implies that the C-APS will defeat the missile with probability unity when it is at a distance of approximately  900 m from $B_1$, when the disruption threshold is $U = 10$.

\subsection{Performance Prediction with Two Missiles}
To illustrate performance for the scenario of two missiles attacking the AFV team it is supposed that the scenario is exactly that in Figure \ref{fig1}. It is assumed that the same characteristics of $M_1(t)$ apply as in the single missile case of the proceeding subsection, so that it is directed towards $B_1$. The second missile, denoted $M_2(t)$ is fired toward $B_2$.
Given the scene geometry, it is assumed that all vehicles provide detection capability for locating missile 1. However, given the proximity of vehicle 1 and 2 to each other, only vehicle 1 cannot contribute to the detection capability for defence of $B_2$. From a missile countering perspective, it is also clear that $B_1$ cannot be used to defend $B_2$. Additionally, $B_4$ may introduce the potential for collateral damage to $B_2$ in an attempt to defeat the second missile. Hence, for this scenario, it will be supposed that $B_1$ and $B_4$ can only be used to counter $M_1(t)$ while $B_2$ and $B_3$ can be used to counter $M_2(t)$. This restriction also avoids the issue of introducing weapon queueing times if a vehicle is able to provide defence against multiple targets.

One can utilise \eqref{R1} - \eqref{R4} as required for the probabilities of detection and disruption of the first missile. Suppose that the second missile travels with a speed of $\nu_2$ and takes $T_2$ to reach $B_2$. Then the distance of the second missile from its target $B_2$ at time $t$ is
\begin{equation}
H_2(t) = \nu_2(T_2 - t). \label{h2}
\end{equation}
Also required are the distances from $M_2(t)$ to $B_3$ and $B_4$, which are denoted $H_3$ and $H_4$ respectively. By an application of the cosine rule it can be shown that
\begin{eqnarray}
H_3(t) &=& \sqrt{2500 + (\nu_2(T_2 -t) + 10)^2 -50 \sqrt{3}( \nu_2(T_2 -t) + 10)} \\\label{h3}
H_4(t) &=& \sqrt{900 + (\nu_2(T_2 -t) + 10)^2 +30 \sqrt{3}( \nu_2(T_2 -t) + 10)}. \label{h4}
\end{eqnarray}
These can then be applied to \eqref{pdf8} and \eqref{mqf1} as required. 

\begin{figure}[h]
	\centering
	\includegraphics[width=13cm]{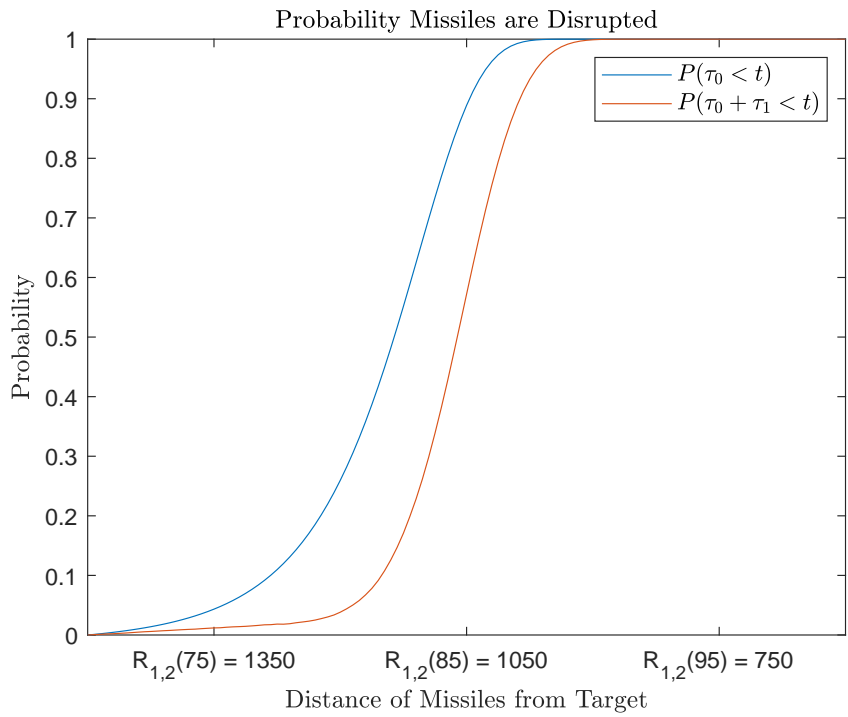}
	\caption{C-APS performance prediction, in the presence of two missiles,  with a common disruption threshold of $U = 10$. $P(\tau_0 < t)$ is the probability one of the two missiles is defeated before reaching distance $R_{1,2}(t)$, while $P(\tau_0 + \tau_1 < t)$ is the probability both missiles are defeated before reaching distance $R_{1,2}(t)$. Both missiles travel to their targets at the same speed and take the same time to intercept their targets. Hence $R_{1, 2}(t)$ is the distance at time $t$ of missile 1 and 2 from their respective targets $B_1$ and $B_2$.}
	\label{fig2a}
\end{figure}

The speed and time to impact of the second missile are assumed to be identical to that of the first missile, and it is also assumed that this missile is identical to the first in terms of its statistical and electromagnetic signature. Figure \ref{fig2a} illustrates the situation where the missile disruption threshold is $U = 10$. The figure plots the probability distribution functions of $\tau_0$ and $\tau_0 + \tau_1$ as functions of time. In view of \eqref{neq2} and \eqref{neq3} the distribution function of $\tau_0$ is plotting the probability that one of the two missiles has been disrupted by time $t$, while that of $\tau_0 + \tau_1$ is plotting the likelihood that both missiles have been defeated. As in the previous figures, the $x$-axis plots the distance of both missiles from their intended targets at time $t$, and so $R_{1, 2}(t)$ is the distance of missile 1 and 2 to its target at time $t$.
From the figure it is clear that both missiles will be defeated within  1 km of their intended targets.

Increasing the threshold $U$ to 100 does not significantly alter these results, but when the threshold is increased to the order of 1000 there is a sharper reduction in performance, as to be expected (figures not included for brevity).

\section{Conclusions and Future Work}
The purpose of this paper was to introduce a continuous time analogue of the discretised AFV defence model developed in \cite{weinberg21}, and to also develop a framework in which HEL DEWs performance could be be assessed. The latter necessitated a novel development of the way in which HEL dwell time is modelled. The model for single missile defence is somewhat simple, while there is a degree in complexity in the two missile case. The examples investigated demonstrated results consistent with expected performance of  HEL DEWs such as the Rheinmetall exemplar discussed previously.

Extensions to three or more missiles is possible but requires more complex conditional probability analysis. Nonetheless, the mathematical model introduced in this paper can be used as a basis for the performance analysis of a team facing multiple missile threats.

\end{document}